# From n-layer planar ordering to the monolayer homeotropic structure of confined hard rods: The effect of shape anisotropy and wall-to-wall separation

Fahimeh Behzadi[1], Seyed Mohammad Ghazi[1], and Roohollah Aliabadi[1,*]

[1]Department of Physics, Faculty of Science, Fasa University, 74617-81189 Fasa, Iran

**Abstract**

Using the Parsons-Lee theory, we examined the effect of shape anisotropy and the wall-to-wall separation ($H$) on the phase behavior of the hard parallelepiped rods with dimensions $L$, $D$, and $D$ ($L > D$) in such narrow slit-like pores which only one homeotropic layer can form. The phase structures, including biaxiality, planar nematic layering transition as well as planar to homeotropic, were studied for some separations in the range $2.5D \leq H \leq 10.0D$ for $H - D \leq L < H$.

**Introduction**

The liquid crystal (LC) phase is still a fascinating subject that has attracted many scientists for decades. Over the years, many scientists have carried out a wide range of investigations on the anisotropic LC molecules experimentally [1, 2], theoretically [3, 4], and using computer simulations [5, 6] due to their widespread applications. Their various properties are widely used in industry, such as sensors [7], photonic instruments [8], and novel bistable devices [9]. The reason for the extensive use of LCDs is the possibility of controlling the LC

*aliabadi@fasau.ac.ir

alignment technologically [10, 11].

Two important groups of LCs are lyotropics and thermotropics. Lyotropics consist of massive particles, and fundamentally their phase transition only depends on their concentration, i.e., the entropy is the driving force [12]. It means the anisotropic shape of the particles is the key factor in the formation of LC phases. Some other factors that affect the phase transition of LC particles are anchoring of the surface, confinements, externally applied electromagnetic fields, and defects [13, 14]. According to the statistics presented in Ref. [15], over the last two decades, lyotropics get even more attention from researchers to understand their fascinating physical properties through theory, simulation, and experimental laboratories. Its modern applications like biotechnology, biosensors, drug delivery, biomimetics, strong nanomembrane films, development of new nanomaterials with new physical properties, and the synthesis of nanoparticles can be named [16-20].

The liquid crystalline states generally may be divided into the isotropic (I), nematic (N), smectic, and columnar phases [4, 21]. These structures can form in both bulk and confined colloidal systems [22,23]. This phase behavior is arising from the strong competition between intrinsic and extrinsic interactions [24]. The confining walls may form a nematic phase with parallel planar (P) or homogenous phase where the long axis of the particle is parallel to the walls and homeotropic structure (H) which the long axis of the particles is perpendicular to the walls. Allen [25, 26] simulated the rigid rod-like particles confined between two planar walls and showed that the preferred anchoring near the walls is planar.

The study of the transition between P and H phases can be of interest from the technological point of view since these phases are essential in designing and constructing optoelectronic devices [11].

The nematic planar phase confined between two walls could be uniaxial (U) or biaxial (B). The U-B transition induced by a substrate(s) has been predicted to be second order for confined hard rod-like [27-30] and plate-like [31] particles. The biaxial nematic phases have been studied widely experimentally [32-34] and theoretically [35-37] due to their fast response time to the applied electric field, which is an essential factor in display technologies [38].

From a practical point of view, studying hard body fluids between two hard walls, as a model system of nanoparticles in nanopores, is also an exciting topic [39, 40] since by decreasing the wall-to-wall separation, it is possible to reach the almost two-dimensional (2D) systems (quasi-2D systems) which the nature of their transition may be completely different with three-dimensional (3D) systems [23]. For example, the I-N phase transition in 3D systems with large enough particles is discontinuous, but in the 2D systems, the N phase changes continuously to the I phase with decreasing density [41, 42].

The ordering of the rectangular hard rod particles in contact with a single wall or between two hard surfaces has been investigated through the theory [27-30, 43]. Their results can be summarized as (i) a wall-induced surface transition from uniaxial to biaxial symmetry, (ii) the nematic film wets the wall completely, (iii) the critical value of the pore width at which the capillary nematization (I-N) terminates is about twice the length of the rod, (iv) the P-H transition occurs at strongly confined nonmesogenic rods (i.e. $1 < L/D < 3$ and $H/D \leq 3$) (v) this first-order transition terminates at about $L/D \approx 2$ when $2 < H/D \leq 3$, (vi) T phase forms for $L/D < 2$, and (vii) planar nematic layering transition takes place for some studied $L/Ds$. Here $L$ is the length of the particle, and $D$ is its short side. Despite some sharp differences, some similar results have also been achieved for hard plate-like particles [31, 44-47].

The understanding and describing the structures of the confined parallelepiped-shaped rods with square or different cross sections are fundamental theoretical problems since such a system is an excellent model for understanding the phase behavior of nanorods like goethite, in nanoconfinement.

In this paper, we examine the effect of very narrow confinement on the surface-induced biaxial ordering and the planar nematic layering transition as well as the transition from P to H phase where $H/D$ is constant, and $L$ is limited to $H - D \leq L < H$ to avoid the formation of complex phases with several mixed structures. It means only one homeotropic layer can place into the pore. Based on our theoretical results, which have been achieved from the Parsons-Lee theory in the restricted orientation approximation, we divide the studied ranges into four parts (i) $2.5 \leq H/D \leq 3.0$, (ii) $3.0 < H/D \leq 4.0$, (iii) $4.0 < H/D \leq 8.0$, and (iv) $H/D \geq 9.0$. First, we study case (iii) where there is one planar nematic layering transition, i.e. $(n - 1)$ layers to n layers. Then we examine the case (iv) or the wider pores where we expect more than one planar nematic layering transition, and in the next step, we calculate the phase borders of the case (ii) where there is no planar nematic layering transition for some pores, but there is $(n - 1) - n$ layering transition for the other pores in this range with presenting re-entrant phenomenon. At last, we study case (i) where there is no planar layering transition.

The drawback of the Parsons-Lee theory is that this theory only captures isotropic and nematic phases, not the solid phase. As a result, the confined n-layer planar fluid may freeze first, and then it transforms into the homeotropic crystal, where both coexisting phases being crystalline. The applied restricted orientation approximation also has quantitative and qualitative effects on the achieved results. To resolve this issue, it is necessary to extend the theory. Therefore, studying the system in a freely rotating manner and including the possible crystalline phases is essential to demonstrate the reported phase structures in this paper. However, the presence of the layering transitions has been proven for confined freely rotating squares between two hard lines in a simulation study [48]. Such a layering phenomenon has also been reported in thin confined films by Monte Carlo Simulation investigations [49,50].

At last, it is worthy to notice that the investigation into the phase structures and surviving the density induced P-H transition in wide pores is essential to fabricate bistable devices or pressure sensors [51].

**Theory**

We studied the confined hard rod particles with a rectangular cross-section and edge lengths $L$, $D$, and $D$ into a slit-like pore with flat and parallel hard walls where the walls are perpendicular to the $z$-axis and placed at $z = 0$ and $z = H$ and spread in the $xy$ plane. We use the Onsager theory with the Parsons-Lee modification [27, 52] within three-state Zwanzig approximation [53], which the orientation freedoms of the particles are along the $x$, $y$, and $z$ directions. Therefore, the local density of each direction ($\rho_i$ where $i = x, y, z$) is a function of the $z$-coordinate only, and we can achieve the packing fraction ($\eta$) from the local densities ($\rho_i(z), i = x, y$ and $z$) as below:

$$\eta = \frac{V_0}{V} \sum_{i=x,y,z} \int d\vec{r} \rho_i(\vec{r}) = \frac{V_0}{H} \sum_{i=x,y,z} \int dz \rho_i(z) \qquad (1)$$

where $V_0 = LD^2$ is the volume of each particle, $L$ is the length of the larger side, and $D$ is the length of shorter sides of a particle, $A$ is the area of the confining surfaces, and $V = AH$ is the volume of the pore. The details of the used theory are in Ref. [27], and here we explained it briefly. It is assumed that the system is athermal, and the free energy only depends on the variation of the entropy ($S$), i.e., the system is entropy-driven. The free energy is the sum of the ideal, excess, and external terms ($F = F_{id} + F_{exc} + F_{ext}$). In the ideal free energy term, the particles' distribution is rather homogeneous in both orientation and position, and it is given by:

$$\frac{\beta F_{id}}{A} = \sum_{i=x,y,z} \int dz \rho_i(z) \left( ln\rho_i(z) - 1 \right) \qquad (2)$$

where $\beta = 1/k_B T$, $T$ is the absolute temperature. The minimized excess free energy term is related to the ordered phases' equilibrium, which is achieved by minimizing the excluded volume between two particles. The second term of the free energy can be written as follows:

$$\frac{\beta F_{\text{exc}}}{A} = \frac{1}{2} c \sum_{i,j=x,y,z} \iint dz_1 dz_2 \rho_i(z_1) \rho_j(z_2) A_{\text{exc}}^{ij}(z_1 - z_2) \qquad (3)$$

where $c = (1 - 3\eta/4)(1 - \eta)^{-2}$ and $A_{\text{exc}}^{ij}$ is the excluded area between two particles with $i$ and $j$ orientations. The external free energy term confines the particles between the hard walls of the pore, and it is written as below:

$$\frac{\beta F_{\text{ext}}}{A} = \sum_{i=x,y,z} \int dz \rho_i(z) \beta V_{\text{ext}}^i(z) \qquad (4)$$

where $V_{\text{ext}}^i(z)$ is zero inside the pore and infinite otherwise. To determine the local equilibrium densities, the free energy is minimized with respect to all density components, and finally, the local equilibrium densities are achieved as below:

$$\rho_k(z) = \frac{H\eta}{V_0} \frac{\exp\left[-\beta V_{\text{ext}}^k(z) - c \sum_{i=x,y,z} \int dz_1 \rho_i(z_1) A_{\text{exc}}^{ik}(z - z_1)\right]}{\sum_{l=x,y,z} \int dz_1 \exp\left[-\beta V_{\text{ext}}^l(z_1) - c \sum_{i=x,y,z} \int dz_2 \rho_i(z_2) A_{\text{exc}}^{il}(z_1 - z_2)\right]} \qquad (5)$$

where $k = x, y, z$. Here, we define the dimensionless distance $z^* = z/D$, the dimensionless density $\rho^* = \rho D^3$ and $H^* = H/D$ where $D$ is unit of the distance.

We use the trapezoidal quadrature rule to solve numerical integrations. The coupled integral equations of the local densities, Eq. (5), have been solved numerically through Picard's iteration method to calculate the equilibrium phase structures. Reliable solutions for each phase depend on choosing the correct initial value. Therefore, it is crucial to find the right initial value of the desired phase at the first packing fraction by an initial function in the code to achieve the solutions at other packing fractions. A linear combination rule is also used to mix the results of the successive iterations. To check the convergence between the results of two consecutive iteration steps, this equation $\frac{1}{1+u} \sum_{i=0}^{u} \sum_k |\rho_k^{j+1}(z_i) - \rho_k^j(z_i)| < 10^{-10}$ is applied where $j$ is an iteration step in which the successive iteration process is considered to be convergent. In addition, we use $\Delta z = 0.01$ where the phase separation is strong and $\Delta z = 0.001$ or $0.0001$ where phase boundaries are very close. Discontinuous phase transitions are placed at the cross point of two different solutions of Eq. (5) in the $\beta\mu - \frac{\beta F}{A}$ plane. It is also important to notice that the stable phase in the system at a certain packing fraction has the smallest free energy.

**Results and Discussion**

We have applied the Parsons-Lee theory to study the phase structures of the hard rectangular rods numerically for $H/D - 1 \leq L/D < H/D$ where only one homeotropic layer can form in the pores, preventing to form complex structures. Therefore, the particles are permitted to lie either planar with maximum n layers or homeotropic with only one layer. The Parsons-Lee formalism is utilized for the prolate [54] and oblate particles [55] and also different types of pores. We use it for rods with aspect ratios in the range of $2.00 \leq L/D <$ $10.00$ with eight different constant wall distances, i.e., $H/D = 2.7$, $2.9$, $3.3$, $3.5$, $4.5$, $6.0$, $8.0$, and $10.0$. For $H/D = 3.3$ and $4.0 < H/D \leq 8.0$, there are only either $(n-1)$ or n planar layers where n is the maximum allowable planar layers, and its value is either n $=$ $[H/D] - 1$ for an integer $H/D$ or n $= [H/D]$ for a non-integer $H/D$ ($[H/D]$ denotes as an integer part of $H/D$). The scenario could be different for other pores. For example, for $H/D =$ $9.0$ and $10.0$, there are two planar layering transitions, i.e. $(n-2) \rightarrow (n-1)$ and $(n-1) \rightarrow$ n (we have not reported the phase diagram of $H/D = 9.0$ ) or for $H/D = 3.5$, and $2.5 \leq$ $H/D \leq 3.0$, there is no planar layering transition.

We present our results as follows: (i) some $H/D$s in the range of $4.0 < H/D \leq 8.0$, (ii) $H/D =$ $10.0$, (iii) some $H/D$s in $3.0 < H/D \leq 4.0$, and (iv) some $H/D$s in $2.5 \leq H/D \leq 3.0$. It would be useful to calculate the highest value of the packing fraction for the planar and homeotropic states to determine the more stable phases. The packing fraction is defined as $\eta =$ $NV_0/V$, where $N$ is the number of particles in the pore. This value for the planar and homeotropic phases are $\eta^P = (NLD/nA) \times (nD/H)$ and $\eta^H = (ND^2/A) \times (L/H)$, respectively, i.e., the packing fraction for each orientation is possible to divide into two dimensional (first parentheses in $\eta^P$ and $\eta^H$) and one dimensional (second parentheses in $\eta^P$ and $\eta^H$) parts. The maximum packing fraction occurs when in each layer the particles' faces are parallel to the walls, and cover the pore surfaces thoroughly, i.e., $(NLD/nA)_{Max} = 1$ for the planar structure and $(ND^2/A)_{Max} = 1$ for the homeotropic one. Therefore, the close packing values of the packing fraction are $\eta_{cp}^P = nD/H$ and $\eta_{cp}^H = L/H$ for the planar and homeotropic structures, respectively. It is worth mentioning that as $\eta_{cp}^P < \eta_{cp}^H < 1$, the homeotropic state will be more stable than the planar phase at high densities. According to the Parsons-Lee theory, the free energy diverges at $\eta = 1$ [56, 57]; therefore, the free energy cannot be infinite at $\eta_{cp}$.

In Fig. 1, we have shown the schematic diagram of the studied phase structures for $H/D = 4.5$, and $L/D = 4.30$. This figure depicts the two-dimensional representation of the possible transitions which occur by increasing $\eta$.

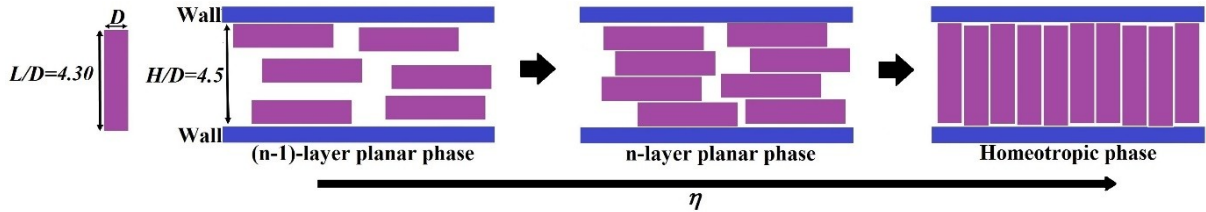

FIG. 1. Schematic diagram of the phase transitions of the particles by increasing $\eta$ for $H/D = 4.5$ and $L/D = 4.30$.

Figure 2(a) presents the phase diagrams of $H/D = 4.5$ with $3.50 \leq L/D < 4.50$. The density profiles of four different packing fractions at $\eta = 0.300, 0.450, 0.550,$ and $0.750$ have been shown in Figs. 2(b), (c), (d), and (e), respectively, which can be identified by different symbols in Fig. 2(a). As we mentioned before, to avoid complicated phases, we set the following condition $H/D - 1 \leq L/D < H/D$. According to Fig. 2(a), by increasing the density, the overall trends of the phase transitions are:

$$3IL \xrightarrow{\text{second order transition}} 3BPL \xrightarrow{\text{first order transition}} 4BPL \xrightarrow{\text{first order transition}} H \text{ (monolayer)},$$ where

3IL, 3BPL, and 4BPL stand for isotropic phase with three layers and planar adsorption at the walls, planar nematic with three biaxial layers, and planar nematic with four biaxial layers, respectively. These behaviors originate from the competition between Eqs. (2) and (3) to increase the available space for the confined rods. Contrary to the free energy's excess term, $F_{\text{exc}}$, the free energy's ideal term, $F_{\text{id}}$, prefers the particles to be in the random positions and orientations. Since the planar ordering is important to decrease the excluded area near the walls, the structure is planar close to the surfaces, as shown in Fig. 2(b). We call this structure the planar isotropic (I) or the uniaxial phase because of the equal contributions of $\rho_x$ and $\rho_y$. By increasing the density, the phase changes to the biaxial ($\rho_x \neq \rho_y \neq \rho_z$) and planar nematic (BP), as illustrated in Fig. 2(c). The (blue) solid line in Fig. 2(a) shows the biaxiality border, which separates the I and the BP phases. This relatively steep slope of the I-BP line arises from the dependency of the second term of the free energy (Eq. (3)) to the aspect ratios of the particles, $L/D$. As shown in Fig. 2(a), the I-BP transition density increases by decreasing $L/D$. In fact, by decreasing the shape anisotropy, the excluded area between two particles decreases [30], and thereby the second term of the free energy decreases, which is compensated by higher $\eta$s. Furthermore, by decreasing $L/D$ in the range of $3.50 \leq L/D < 4.50$, the aspect ratio is getting closer to the region of nonmesogenic particles ($L/D < 3.00$), and the I-BP transition takes place even at higher packing fractions for more isotropic particles [27]. The slope of all studied I-BP lines are presented in Table 1. According to this table, the absolute value of the slope of all the reported U-B lines is decreasing by increasing the pore width as we expect due to the increment of the shape anisotropy of the studied particles with increasing the wall-to-wall distance.

As illustrated in Fig. 2(a), (b), and (c) both the isotropic and the biaxial phases have three layers. There is no 2-3 planar layering transition for $H/D = 4.5$ since the pore is wide enough to form three planar layers mixing with out-of-plane particles even at the low densities.

By comparing Figs. 2(c) and (d), it is clear that by increasing the density, the first order 3BPL-4BPL transition takes place where its coexisting region is patterned by tilting lines and green color in Fig. 2(a). The creation of a new planar layer is discontinuous since it pushes the existing layers to the direction of the walls, and thereby there is less space for old layers. The dashed (violet) line shows the close packing value of the packing fraction of the 3BPL phase, $\eta_{cp}^{3BPL} = 0.667$. The occurrence of the 3BPL-4BPL transition below $\eta_{cp}^{3BPL}$ shows the high stability of the 4BPL structure. The 3BPL-4BPL transition depends very weakly on $L/D$ and decreases very slowly by the decrement of $L/D$. This dependency arises from the easier rearrangement of the particles with smaller aspect ratios.

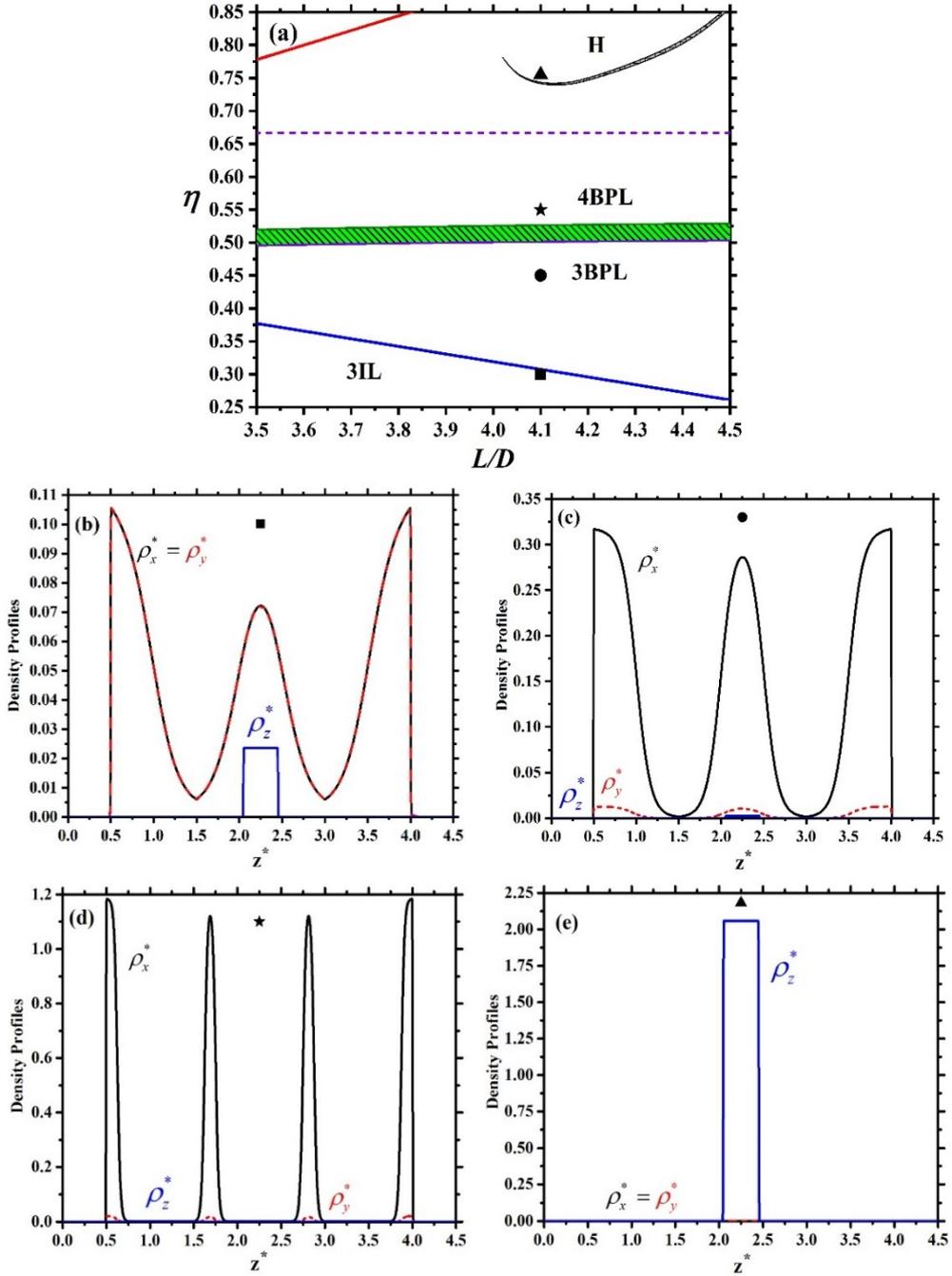

FIG. 2. (a) The coexisting packing fractions versus $L/D$ for $H/D = 4.5$, (b)-(e) the density profiles of the different phases versus $z^*$ for $L/D = 4.10$ marked by different symbols in (a). These phases are (b) planar isotropic with three layers (3IL) at $\eta = 0.300$, (c) biaxial planar nematic with three layers (3BPL) at $\eta = 0.450$, (d) biaxial planar nematic with four layers (4BPL) at $\eta = 0.550$, and (e) homeotropic (H) at $\eta = 0.750$.

As it is apparent in Fig 2(a), by increasing the packing fraction, the first-order 4BPL-H phase transition takes place (grey region) for some $L/D$s. This change in the structure can also be seen by comparing Fig. 2(d) and (e). Since the available distances along the z-axis in the planar and homeotropic phases are $(H - D)$ and $(H - L)$, respectively, the inequality $H - D > H - L$ guarantees more available room for the planar phase for the prolate particles $(L/D > 1)$. However, by increasing the density, the particle-particle interactions increase, and the particles distribute themselves in such a way that the free energy of the system decreases, thereby causing the homeotropic phase. This P-H transition takes place in the range of $4.02 \leq L/D < 4.50$ where the minimum $\eta$ or the 4BPL-H transition curve corresponds to $L/D = 4.13$, and it goes up to the higher values of $\eta$ as $L/D \to 4.50$ and $4.02$. This transition terminates at $(L/D)_c \approx 4.02$, and $\eta_c \approx 0.785$. It means that for $L/D < 4.02$, 4BPL phase evolves continuously into the H phase on increasing $\eta$ because the rods with smaller elongation can rearrange easier, hence it is easier to accommodate the homeotropic layer into the pore.

The solid (red) line in the top-left of Fig. 2(a) shows $\eta_{\text{cp}}^{\text{H}}$, which grows by increasing $L/D$. As shown in this figure, the H phase happens below $\eta_{\text{cp}}^{\text{4BPL}} = 0.889$ (not shown in the figure) which indicates the stability of this structure. Since in the homeotropic phase, by increasing the particles' aspect ratios, the translational entropy decreases along the z-axis, the system prefers to remain in the planar phase rather than the homeotropic one. That's why the 4BPL phase for $4.13 \leq L/D < 4.50$ has the widening stability region. Note that the stability of the 4BPL phase with respect to the homeotropic phase for $4.02 \leq L/D < 4.13$ cannot be explained by the same reason because, in the homeotropic order, the translational entropy is now larger. Therefore, the main factor in this phenomenon could be the packing entropy as the close packing values of 4BPL and H phases are becoming closer to each other as $L/D$ goes to $4.02$.

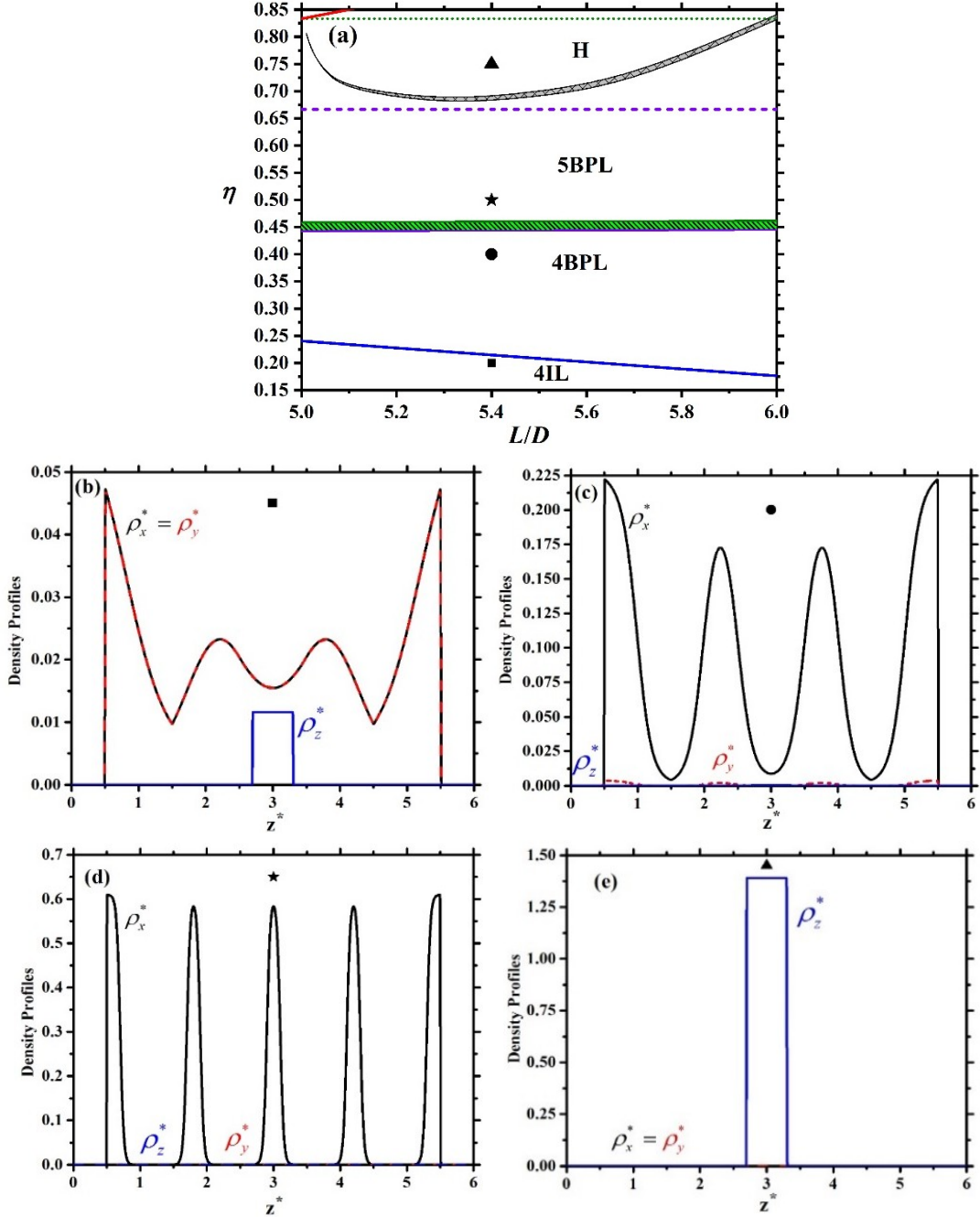

FIG. 3. (a) The coexisting packing fractions versus $L/D$ for $H/D = 6.0$, (b)-(e) the density profiles of different phases versus $z^*$ for $L/D = 5.40$ marked by different symbols in (a). These phases are (b) planar isotropic with four layers (4IL) at $\eta = 0.200$, (c) biaxial planar nematic with four layers (4BPL) at $\eta = 0.400$, (d) biaxial planar nematic with five layers (5BPL) at $\eta = 0.500$ , and (e) homeotropic (H) at $\eta = 0.750$.

Figure 3 shows the results of $H/D = 6.0$ when $5.00 \leq L/D < 6.00$. The different phase structures for $H/D = 6.0$ are indicated in Fig. 3(a). Figures 3(b) and (c) depict the structure changes from a planar isotropic with four layers (4IL) to the biaxial planar nematic with the same number of layers (4BPL). Although there exist particles with homeotropic order in the middle of the pore in the 4IL phase, by increasing the density, the fraction of the particles with the homeotropic orientation decreases (Fig. 3(b)) because of their interaction with all planar layers. Therefore, the planar ordering wins the competition, and more planar layers emerge in the pores, i.e., 4BPL changes to 5BPL (Figs. 3(c) and (d)). The maximum number of planar layers is 5 for $H/D = 6.0$, and the discontinuous transition from 4BPL to 5BPL happens around $\eta = 0.450$. The $\eta_{\mathrm{cp}}^{\mathrm{4BPL}}$ and $\eta_{\mathrm{cp}}^{\mathrm{5BPL}}$ are 0.667 and 0.833, respectively, as shown by a dashed violet and a dotted green lines in Fig. 3(a). However, at very high densities (homeotropic ordering), the packing entropy (the excluded area) term of the free energy $(F_{\mathrm{exc}})$ wins over the translational and orientational ones ($\rho_i \ln \rho_i$) due to minimizing the excluded area term between the particles in Eq. (3), thereby 5BPL-H transition takes place. The gray area with rhombus pattern is the coexisting region of 5BPL and homeotropic phases; and the solid (red) line on the left-top of Fig. 3(a) shows $\eta_{\mathrm{cp}}^{\mathrm{H}}$. This first-order transition terminates at $(L/D)_c \approx 5.01$ and $\eta_c \approx 0.807$. The slope of the I-BP transition line in Fig. 3(a) is smaller than the one in Fig. 2(a) and occurs at lower densities due to the more anisotropic shape of the particles (Table 1).

Similar transitions with different number of layers can also be seen in Fig. 4 for $H/D = 8.0$.

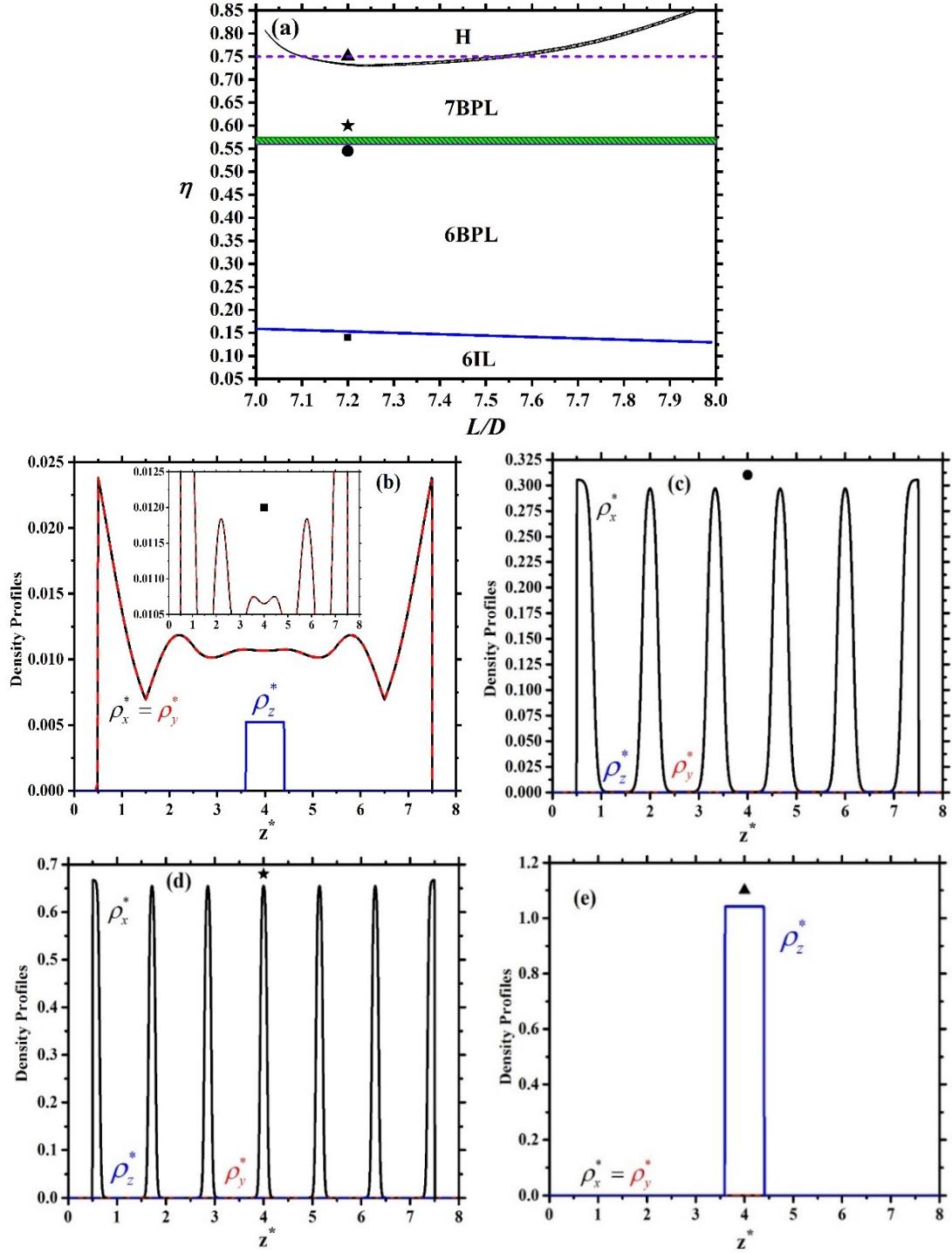

FIG. 4. The phase structures of confined hard rods for $H/D = 8.0$, (a) The coexisting packing fractions versus $L/D$, (b)-(e) the density profiles of different phases versus $z*$ for $L/D = 7.20$ marked by different symbols in (a). These phases are (b) planar isotropic with six layers (6IL) at $\eta = 0.140$. The inset shows part of in-plane density profiles, and it depicts the number of planar layers clearly, (c) biaxial planar nematic with six layers (6BPL) at $\eta = 0.550$, (d) biaxial planar nematic with seven layers (7BPL) at $\eta = 0.600$, and (e) homeotropic (H) at $\eta = 0.750$.

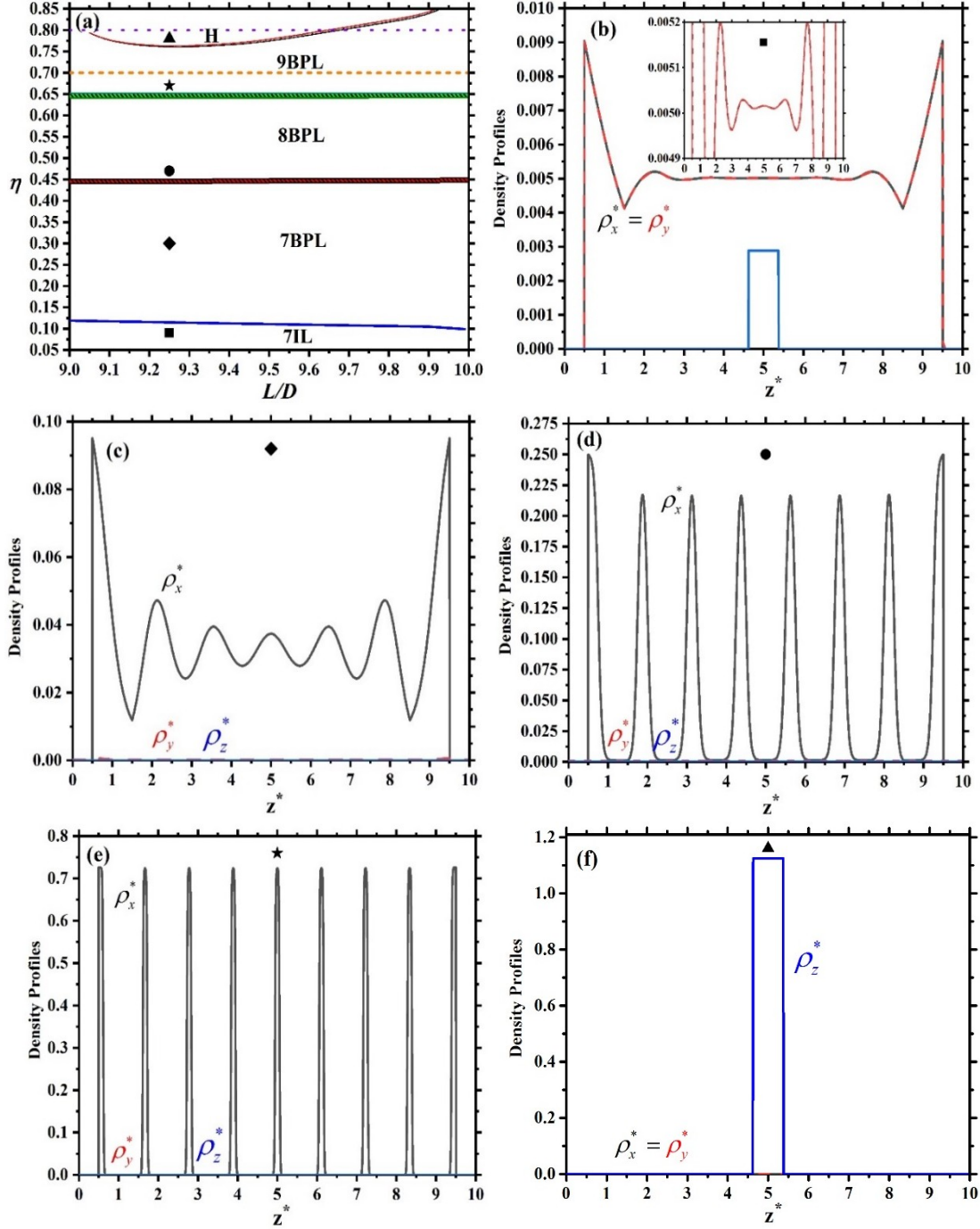

FIG. 5. The phase structures of confined hard rods for $H/D = 10.0$ (a) The coexisting packing fractions versus $L/D$ , (b)-(f) the density profiles of different phases versus $z^*$ for $L/D = 9.25$ marked by different symbols in (a). These phases are (b) planar isotropic with seven layers (7IL) at $\eta = 0.090$. The inset shows part of in-plane density profiles, and it depicts the number of planar layers clearly, (c) biaxial planar nematic with seven layers (7BPL) at $\eta = 0.300$, (d) biaxial planar nematic with eight layers (8BPL) at $\eta = 0.470$, (e) biaxial planar nematic with nine layers (9BPL) at $\eta = 0.670$, and (f) homeotropic (H) at $\eta = 0.780$.

As it is obvious from this figure, the critical point of 7BPL-H transition occurs at $(L/D)_c \approx 7.02$ and $\eta_c \approx 0.806$. The dashed (violet) line shows $\eta_{cp}^{6BPL} = 0.750$ and the inset in Fig. 4(b) presents the number of layers clearly. The slope of the (blue) line in Fig. 4(a) is smaller than the ones in Figs. 2(a) and 3(a) and occurs at lower densities due to the more anisotropic shape of the particles (Table 1).

For all the studied $H/D$s in the range of $4.0 < H/D \leq 8.0$, only $(n-1)$ to n layering transition occurs, but more layering transitions emerge for studied wider pores. In this order, we examined the layering transitions of $H/D = 9.0$ and $10.0$. We report the phase structures of $H/D = 10.0$ in Fig. 5 where in addition to 8BPL-9BPL transition, the 7BPL-8BPL transition also occurs. According to our results, by increasing the wall-to-wall distance and the shape anisotropy, we expect the occurrence of more and more layering transitions for $H/D \geq 9.0$ because in the middle of the pore, the local density is almost constant and the new peak can emerge easily in that area. In Fig. 5(a), the dashed (orange) line, and the dotted (violet) one show the $\eta_{cp}^{7BPL} = 0.700$ and $\eta_{cp}^{8BPL} = 0.800$, respectively. The critical point of the 9BPL-H transition occurs at $(L/D)_c \approx 9.05$ and $\eta_c \approx 0.794$. The inset in Fig. 5(b) enables us to count the number of the existing layers.

In this paper, we have also studied the phase structures of some pores for $3.0 < H/D \leq 4.0$ and $2.5 \leq H/D \leq 3.0$.

Figure 6(a) and (b) show the phase diagrams of $H/D = 3.3$ and $H/D = 3.5$, respectively. As shown in this figure, the 2IL-2BPL transition line (blue solid line in Fig. 6(a)) intersects 2BPL-3BPL coexisting region at $L/D = 2.89$, and at higher densities, 3IL-3BPL transition emerges and intersects this region at $L/D = 3.05$ (red dotted line in Fig. 6(a)). Therefore, the phase structures at the intermediate densities for $H/D = 3.3$, could be divided into three parts depending on the $L/D$ values. By increasing $\eta$, the phase transitions are as follows:

1: 2IL-2BPL-3BPL for $3.05 \leq L/D < 3.30$

2: 2IL-2BPL-3IL-3BPL for $2.89 \leq L/D < 3.05$

3: 2IL-3IL-3BPL for $2.85 \leq L/D < 2.89$

Note that the re-entrant phenomenon takes place in part 2, i.e., $2.89 \leq L/D < 3.05$. Here both the isotropic and biaxial planar phases are re-entrant. The isotropic phase with two planar layers (at lower densities) and the isotropic phase with three planar layers (at higher densities) change to two and three biaxial planar layers, respectively. Four different density profiles of $H/D = 3.3$ can be seen in Figs. 6(c), (d), (e) and (f). For $L/D < 2.89$, there exists a direct 2IL-3IL transition, which causes three planar isotropic layers ordering emerges without accruing 2BPL phase in this region. Figs. 6(e) and (f) show that the peaks are narrower for three-layer states than the peaks of the planar isotropic with two layers (Figs. 6(c) and (d)), which means that for 3IL and 3BPL, the particles near the opposite walls do not overlap with each other. Since the particles are packed very well with three-layer uniaxial planar state in the pore as well as being in the in-bulk nonmesogenic particle range ($L/D < 3.00$), make it possible to have the planar isotropic ordering at higher densities. As a result, the second-order 3IL-3BPL transition shifts to higher densities. The close packing value of two layers (the violet dashed line in Fig. 6(a)) and three layers (not shown in Fig. 6(a)) are $\eta_{\mathrm{cp}}^{\mathrm{2BPL}} = 0.606$ and $\eta_{\mathrm{cp}}^{\mathrm{3BPL}} = 0.909$, respectively. The phase changes from 3BPL to H by increasing the density and this transition terminates at $(L/D)_c \approx 3.01$ and $\eta_c \approx 0.777$. Figure 6(b) depicts the phase structures of $H/D = 3.5$. There is no layering transition for $H/D = 3.5$ and hence $3.5 < H/D \leq 4.0$ since in this range the pore is wide enough to form three planar layers even at low densities but not wide enough to have four planar layers. At higher values of the densities the phase transition of 3BPL to H is formed which terminates at $(L/D)_c \approx 3.01$ and $\eta_c \approx 0.784$. The close packing value of three-layer state is $\eta_{\mathrm{cp}}^{\mathrm{3BPL}} = 0.857$ that has not been shown in Fig. 6(b).

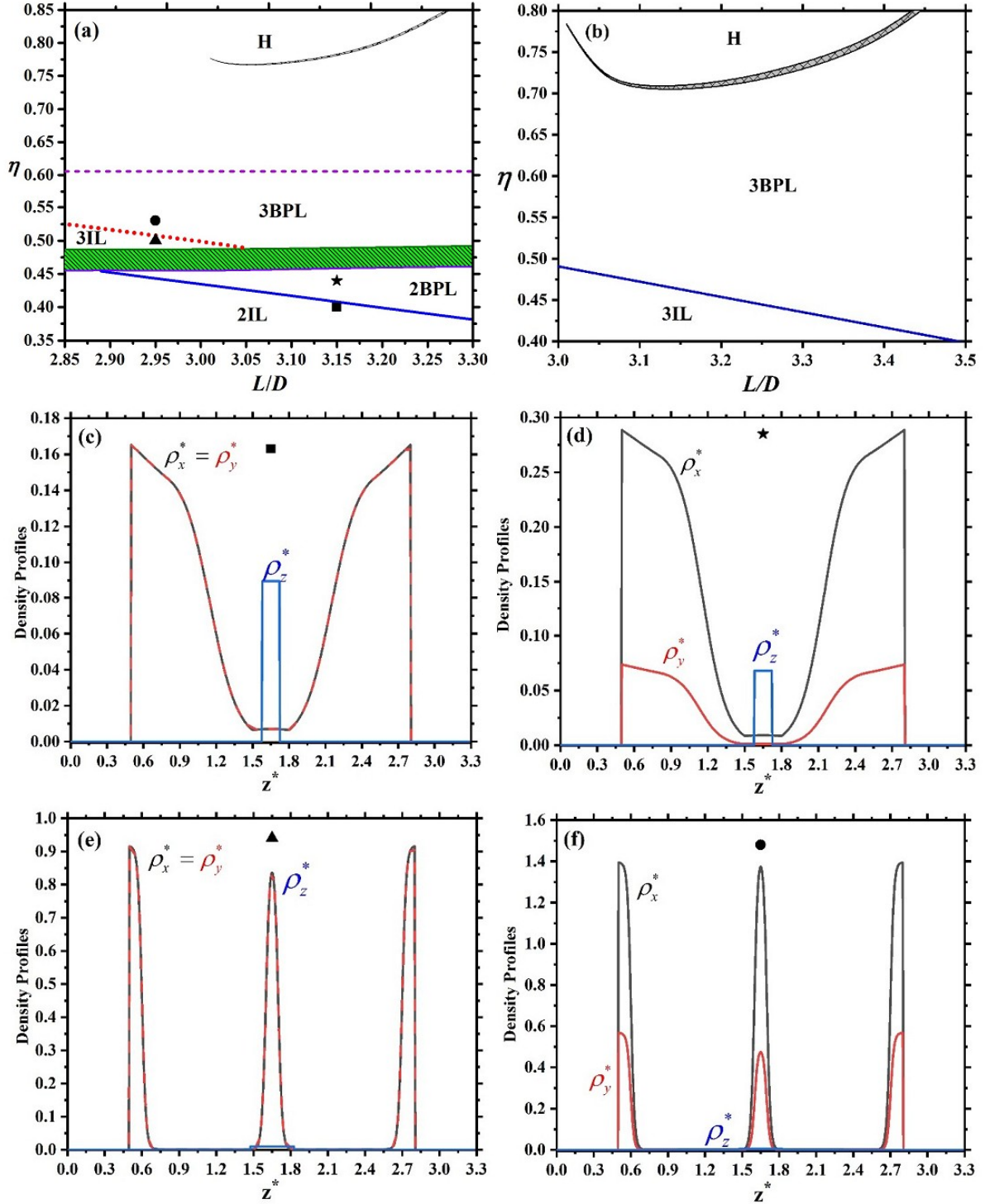

FIG. 6. The phase structures of confined hard rods for (a) $H/D = 3.3$, (b) $H/D = 3.5$. (c)-(f) The density profiles of different phases of $H/D = 3.3$ versus $z^*$ marked by different symbols in (a). These phases are (c) planar isotropic with two layers (2IL) at $\eta = 0.400$ and $L/D = 3.15$, (d) biaxial planar nematic with two layers (2BPL) at $\eta = 0.440$ and $L/D = 3.15$, (e) planar isotropic with three layers (3IL) at $\eta = 0.500$ and $L/D = 2.95$, and (f) biaxial planar nematic with three layers (3BPL) at $\eta = 0.530$ and $L/D = 2.95$.

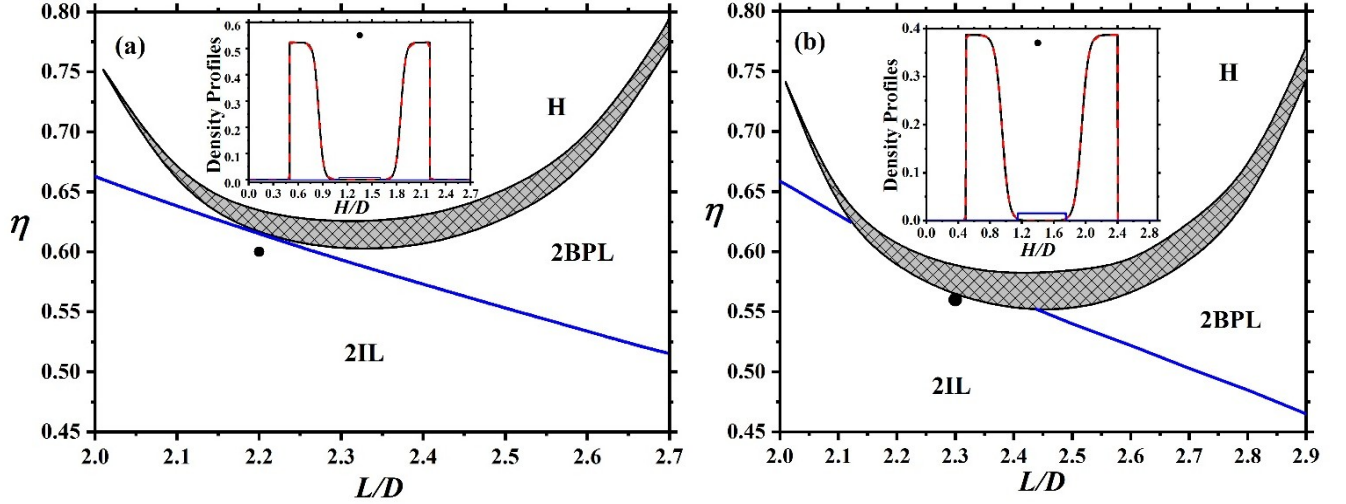

FIG. 7. The phase transitions of confined hard rods for (a) $H/D = 2.7$ with the density profiles of the planar isotropic (in the inset) with two layers (2IL) at $\eta = 0.600$ and $L/D = 2.20$ and (b) $H/D = 2.9$ with the density profiles of the planar isotropic (in the inset) with two layers (2IL) at $\eta = 0.560$ and $L/D = 2.30$.

According to Ref. [27], there exists a direct transition from I to H phase for $2.10 < L/D <$ 2.50 when $H/D = 3.0$, i.e., the BP phase is absent in above mentioned range, and this transition does not happen for $H/D \leq 2.5$. To investigate if this direct transition exists for the other $H/D$s in the range of $2.5 < H/D < 3.0$, we calculated the phase separations of $H/D$s in this range and reported the phase diagrams of $H/D = 2.7$ and 2.9 as shown in Figs. 7(a) and (b), respectively. Considering the findings of Ref. [27] and our calculations, we conclude that there is a direct I-H transition in the range of $2.7 \leq H/D \leq 3.0$. The insets in Fig. 7(a) and (b) show the density profiles at $\eta$s which are below the I-H coexisting region. The 2BPL-H transition terminates at $(L/D)_c \approx 2.01$ and $\eta_c \approx 0.751$ for $H/D = 2.7$, and $(L/D)_c \approx 2.01$ and $\eta_c \approx 0.740$ for $H/D = 2.9$. As it is clear, for $H/D \leq 3.0$, the homeotropic phase appears at lower densities with respect to $H/D > 3.0$. Because for wider pores, the translational entropy is more dominant and prevents forming the homeotropic ordering at lower densities. For $H/D \leq 2.5$, the pore widths are very narrow; therefore, switching from the planar phase to the homeotropic one is harder, and the P-H transition takes place at higher densities than the densities that the I-BP transition occurs [27], but the pores in the range of $2.7 \leq H/D \leq 3.0$ are wide enough to change the phase from planar to homeotropic at the densities where the I-BP transition can form.

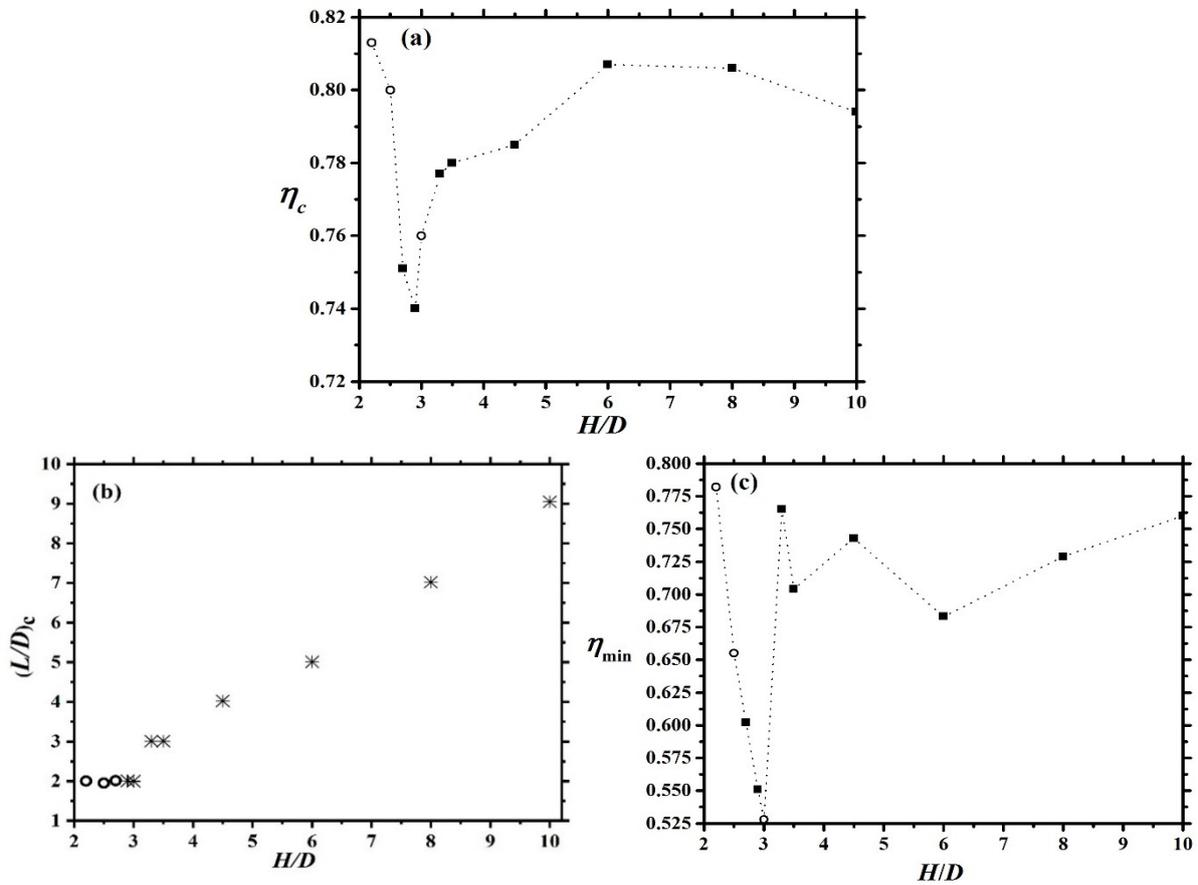

FIG. 8. (a) The critical packing fractions (b) the critical aspect ratios of the nBPL-H transitions and (c) the minimum coexisting packing fractions in nBPL-H transitions versus wall-to-wall separation. The hollow circles are related to Ref. [27] and other symbols are related to our calculations.

The critical packing fractions of the critical points of nBPL-H transitions versus $H/D$ have been shown in Fig. 8(a). The hollow circles are related to Ref. [27], and the filled squares belong to the calculations of this paper with dotted line as a guide to the eye. As it can be seen, the lowest $\eta_c$ of the studied cases belongs to $H/D = 2.9$, and the highest $\eta_c$s are for $H/D = 2.2, 6.0$ and $8.0$. The critical aspect ratio of the particles ($(L/D)_c$) versus wall-to-wall separation is illustrated in Fig. 8(b). The hollow circles are corresponded to Ref. [27], and the asterisks belong to our calculations. According to this figure, for all the pores where $H_0/D <$ $H/D \leq H_0/D + 1$, which $H_0/D$ is an integer number, the critical point of the nBPL-H transitions occur approximately around $(L/D)_c \approx H_0/D$. However, to express a general rule, more wall-to-wall distances have to be studied. Fig. 8(c) presents the lowest coexisting densities in nBPL-H transitions versus $H/D$s. Figure 8 depicts that the best $H/D$ range to investigate nBPL-H transition for confined rectangular rods in future simulation and experimental studies is about $2.7 \leq H/D \leq 3.0$ at $\frac{L}{D} \approx 2 + \frac{\frac{H}{D}-2}{2}$ as well as $\frac{H}{D} = 6.0$ at $\frac{L}{D} \approx 5.3$. According to Fig. 8(c) for $2.2 \leq H/D \leq 3.0$, by increasing the pore width, the H phase could be stabilized at lower densities since there is no planar layering transition in this range, and by increasing the wall separation the particles can order easier at the direction normal to the walls. At larger pore widths (i.e. $\frac{H}{D} > 3.0$) the planar layering transitions shift the nBPL-H transition to higher densities.

Table 1. The slope of the I-BP lines of different $H/D$s. The values marked with * have been reported in Ref. [27].

| $H/D$ | $L/D$ | The range of $\eta$ for I-BP transition | The slope of I-BP |
|---|---|---|---|
| 2.2* (2IL-2BPL) | $1.60 \leq L/D < 2.20$ | $0.800 \geq \eta \geq 0.648$ | -0.253 |
| 2.5* (2IL-2BPL) | $1.74 \leq L/D < 2.50$ | $0.750 \geq \eta \geq 0.561$ | -0.236 |
| 2.7 (2IL-2BPL) | $2.00 \leq L/D \leq 2.69$ | $0.663 \geq \eta \geq 0.516$ | -0.213 |
| 2.9 (2IL-2BPL) | $2.00 \leq L/D \leq 2.89$ | $0.657 \geq \eta \geq 0.468$ | -0.212 |
| 3.0* (2IL-2BPL) | $1.95 \leq L/D < 3.00$ | $0.658 \geq \eta \geq 0.443$ | -0.205 |
| 3.3 (3IL-3BPL) | $2.85 \leq L/D \leq 3.05$ | $0.525 \geq \eta \geq 0.489$ | -0.193 |
| 3.3 (2IL-2BPL) | $2.89 \leq L/D \leq 3.29$ | $0.455 \geq \eta \geq 0.380$ | -0.177 |
| 3.5 (3IL-3BPL) | $3.00 \leq L/D \leq 3.49$ | $0.491 \geq \eta \geq 0.407$ | -0.169 |
| 4.5 (3IL-3BPL) | $3.50 \leq L/D \leq 4.49$ | $0.377 \geq \eta \geq 0.262$ | -0.116 |
| 6.0 (4IL-4BPL) | $5.00 \leq L/D \leq 5.99$ | $0.240 \geq \eta \geq 0.177$ | -0.064 |
| 8.0 (6IL-6BPL) | $7.00 \leq L/D \leq 7.99$ | $0.159 \geq \eta \geq 0.130$ | -0.029 |
| 10.0 (7IL-7BPL) | $9.00 \leq L/D \leq 9.99$ | $0.119 \geq \eta \geq 0.099$ | -0.020 |

**Conclusion**

We have investigated the effect of the pore width and aspect ratio on the phase behaviors of hard rod-like particles located between two parallel hard walls. Through the Parsons-Lee theory in a restricted orientation model. We have found that the purely repulsive forces between walls and particles induce a biaxial nematic order with strong adsorption at the walls where the nematic director is parallel to the walls. This strong surface adsorption maximizes the available space for the particles in the pore, reducing the excluded volume cost between the particles. It is demonstrated that if the surface density of adsorbed particles at the walls exceeds the I-N transition density of 2D hard rectangles, a surface induces the biaxial nematic phase. The biaxial ordering occurs at higher densities by decreasing shape anisotropy due to the less packing entropy gain with in-plane ordering. We observed the biaxiality in all the studied wall-to-wall distances. However, for $2.7 \leq H/D \leq 3.0$ and some $L/D$s, the H phase overcomes the biaxiality, and for some $H/D$s between 3.0 and 4.0, two types of biaxiality occur.

Since the pores are not wide enough in the middle of the pore, the inhomogeneous fluid structure cannot relax the same as the bulk, and the wall effects determine the number of layers. We have observed that there are always two planar layers at the walls, but a homeotropically ordered layer competes with the planar one in the middle of the pore. For $3.0 < H/D \leq 4.0$, layering transition occurs for some $H/D$s with re-entrant phenomenon but for other $H/D$s in this range; there is no planar layering transition. For $4.0 < H/D \leq 8.0$, only $(n - 1)$ to n layering transition occurs, and finally, for $H/D \geq 9.0$, there is $(n - 2)$BPL-$(n - 1)$BPL-nBPL-H phase sequence with increasing the density. Note that for all studied $H/D$s in this paper, the $(n - 1)$BPL-nBPL transitions occur below $\eta_{cp}^{(n-1)BPL}$, which shows the high stability of nBPL phases. By comparing the studied pore widths, it is obvious that the width of the coexisting region of $(n - 1)$BPL-nBPL transition is smaller for wider pores.

In 2018, the planar-to-homeotropic transition has been reported for $H/D \leq 3.0$ [27] and here we show this kind of transition survives even for wider pores. For all studied $H/Ds$ in this paper, the nBPL-H transitions take place at $\eta < \eta_{cp}^{nBPL}$, which proves the stability of the H phase. Such a transition has been reported for stiff-polymer rings [58] and also examined by changing the wall penetrability in a system of hard Gaussian overlap particles [59].

. **ACKNOWLEDGMENTS**

R. A. is indebted to Prof. Szabolcs Varga (University of Pannonia, Hungary) for his generosity and permanent scientific supports. Authors also thank Fasa University for providing supercomputing facilities.

**References**

[1]    M. Okumuş, H. Eskalen, M. Sünkür, and Ş. Özgan, Mesogenic properties of PAA/6BA binary liquid crystal complexes, J. Mol. Struct. **1178**, 428 (2019).

[2]    M. Yao et al., Synthesis and properties of new non-symmetric liquid crystal dimers containing mandelic acid and cyano group, Liq. Cryst. **45**, 931 (2018).

[3]    R. Aliabadi, M. Moradi, and S. Varga, Tracking three-phase coexistences in binary mixtures of hard plates and spheres, J. Chem. Phys. **144**, 074902 (2016).

[4]    L. Mederos, E. Velasco, and Y. Martínez-Ratón, Hard-body models of bulk liquid crystals, J. Phys. Condens. Matter **26**, 463101 (2014).

[5]    C. M. Care, and D. J. Cleaver, Computer simulation of liquid crystals, Rep. Prog. Phys. **68**, 2665 (2005).

[6]    M. Aghaei Semiromi and A. Avazpour, Anchoring transition of confined prolate hard spherocylinder liquid crystals: Hard needle-wall potential, Liq. Cryst. **45**, 1396 (2018).

[7]    Ch. Luan, H. Luan, and D. Luo, Application and technique of liquid crystal-based

biosensors, Micromachines **11**, 176 (2020).

[8]   R. Manda et al., Self-supported liquid crystal film for flexible display and photonic
      applications, J. Mol. Liq. **291**, 111314 (2019).

[9]   X. Du, Y. Li, Y. Liu, F. Wang, and D. Luo, Electrically switchable bistable dual
      frequency liquid crystal light shutter with hyper-reflection in near infrared, Liq. Cryst.
      **46**, 1727 (2019).

[10]  H. W. Chen, J. H. Lee, B. Y. Lin, S. Chen, and S. T. Wu, Liquid crystal display and
      organic light-emitting diode display: Present status and future perspectives, Light: Sci.
      Appl. **7**, 17168 (2018).

[11]  V. Kumar et al., Highly stable, pretilted homeotropic alignment of liquid crystals
      enabled by in situ self-assembled, dual-wavelength photoalignment, ACS Appl.
      Electron. Mater. **2**, 2017 (2020).

[12]  L. Onsager, The effects of shape on the intraction of colloidal particles, Ann. Acad.
      Sci. **51**, 627 (1949).

[13]  F. C. Frank, Liquid crystals: On the theory of liquid crystals, Discuss. Faraday Soc. **25**,
      19 (1958).

[14]  P. G. de Gennes, J. Prost, *The physics of liquid crystals* (Clarendon Press, Oxford,
      1993) 2$^{nd}$ ed.

[15]  I. Dierking, and A. M. F. Neto, Novel trends in lyotropic liquid crystals, Crystals **10**,
      604 (2020).

[16]  S. V. Shiyanovskii et al., Lyotropic chromonic liquid crystals for biological sensing
      applications, Mol. Cryst. Liq. Cryst. **434**, 259 (2005).

[17]  E. Oton, J. M. Oton, M. Caño-García, J. M. Escolano, X. Quintana, and M. A. Geday,
      Rapid detection of pathogens using lyotropic liquid crystals, Opt. Express **27**, 10098
      (2019).

[18]   J. F. De Souza et al., Spotlight on biomimetic systems based on lyotropic liquid crystal, Molecules **22**, 419 (2017).

[19]   T. M. Dellinger, and P. V. Braun, Lyotropic liquid crystals as nanoreactors for nanoparticle synthesis, Chem. Mater. **16**, 2201 (2004).

[20]   B. J. Coscia, J. Yelk, M. A. Glaser, D. L. Gin, X. Feng, and M. R. Shirts, Understanding the nanoscale structure of inverted hexagonal phase lyotropic liquid crystal polymer membranes, J. Phys. Chem. B **123**, 289  (2018).

[21]   L. B. G. Cortes, Y. Gao, R. P. A. Dullens, and D. G. A. L. Aarts, Colloidal liquid crystals in square confinement: Isotropic, nematic and smectic phases, J. Phys.: Condens. Matter **29**, 064003 (2017).

[22]   G. J. Vroege and H. N. W. Lekkerkerker, Phase transitions in lyotropic colloidal and polymer liquid crystals, Rep. Prog. Phys. **55**, 1241 (1992).

[23]   E. Basurto, P. Gurin, S. Varga, and G. Odriozola, Ordering, clustering, and wetting of hard rods in extreme confinement,  Phys. Rev. Research **2**, 013356 (2020).

[24]   D. de las Heras, E. Velasco, and L. Mederos, Capillary effects in a confined smectic phase of hard spherocylinders: Influence of particle elongation, Phys. Rev. E **74**, 011709 (2006).

[25]   M. P. Allen, Molecular simulation of liquid crystals, Mol. Phys. **117**, 2391 (2019).

[26]   M.P. Allen, Molecular simulation and theory of the isotropic–nematic interface, J. Chem. Phys. **112**, 5447 (2000).

[27]   R. Aliabadi, P. Gurin, E. Velasco, and S. Varga, Ordering transitions of weakly anisotropic hard rods in narrow slitlike pores, Phys. Rev. E **97**, 012703 (2018).

[28]   R. van Roij, M. Dijkstra, and R. Evans, Interfaces, wetting, and capillary nematization of a hard-rod fluid: Theory for the Zwanzig model, J. Chem. Phys. **113**, 7689 (2000).

[29]   R. van Roij, M. Dijkstra, and R. Evans, Orientational wetting and capillary

nematization of hard-rod fluids, Europhys. Lett. **49**, 350 (2000).

[30]    R. Aliabadi, M. Moradi, and S. Varga, Orientational ordering of confined hard rods: The effect of shape anisotropy on surface ordering and capillary nematization, Phys. Rev. E **92**, 032503 (2015).

[31]    S. Mizani, R. Aliabadi, H. Salehi, and S. Varga, Orientational ordering and layering of hard plates in narrow slitlike pores, Phys. Rev. E **100**, 032704 ( 2019 ).

[32]    G. R. Luckhurst, Biaxial nematic liquid crystals: fact or fiction? Thin solid films **393**, 40 (2001).

[33]    L. A. Madsen, T. J. Dingemans, M. Nakata, and E. T. Samulski, Thermotropic biaxial nematic liquid crystals, Phys. Rev. Lett. **92**, 145505 (2004).

[34]    H. Mundoor, S. Park, B. Senyuk, H. H. Wensink, and I. I. Smalyukh, Hybrid molecular-colloidal liquid crystals, Science **360**, 768 (2018).

[35]    R. Alben, Liquid crystal phase transitions in mixtures of rodlike and platelike molecules, J. Chem. Phys. **59**, 4299 (1973).

[36]    R. Berardi, and C. Zannoni, Do thermotropic biaxial nematics exist? A Monte Carlo study of biaxial Gay–Berne particles, J. Chem. Phys. **113**, 5971 (2000).

[37]    R. A. Skutnik, L. Lehmann, S. Püschel-Schlotthauer, G. Jackson, and M. Schoen, The formation of biaxial nematic phases in binary mixtures of thermotropic liquid-crystals composed of uniaxial molecules, Mol. Phys. **117**, 2830 (2019).

[38]    J. C. Eichler, R. A. Skutnik, A. Sengupta, M. G. Mazza, and M. Schoen, Emergent biaxiality in nematic microflows illuminated by a laser beam, Mol. Phys. **117**, 3715 (2019).

[39]    H. H. Wensink et al., Differently shaped hard body colloids in confinement: From passive to active particles, Eur. Phys. J.: Spec. Top. **222**, 3023 (2013).

[40]    A. B. G. M. Leferink op Reininka, E. van den Pol, A. V. Petukhov, G. J. Vroege, and

H. N. W. Lekkerkerker, Phase behaviour of lyotropic liquid crystals in external fields and confinement, Eur. Phys. J. Spec. Top. **222**, 3053 (2013).

[41]  J. M. Kosterlitz, and D. J. Thouless, Ordering, metastability and phase transitions in two-dimensional systems, J. Phys. C: Solid State Phys. **6**, 1181 (1973).

[42]  D. Frenkel, and R. Eppenga, Evidence for algebraic orientational order in a two-dimensional hard-core nematic, Phys. Rev. A **31**, 1776 (1985).

[43]  H. Salehi, S. Mizani, R. Aliabadi, and S. Varga, Biaxial layering transition of hard rodlike particles in narrow slitlike pores, Phys. Rev. E **98**, 032703 (2018).

[44]  L. Harnau, and S. Dietrich, Fluids of platelike particles near a hard wall, Phys. Rev. E **65**, 021505 (2002).

[45]  L. Harnau, and S. Dietrich, Wetting and capillary nematization of binary hard-platelet and hard-rod fluids, Phys. Rev. E **66**, 051702 (2002).

[46]  D. van der Beek et al., Isotropic-nematic interface and wetting in suspensions of colloidal platelets, Phys. Rev. Lett. **97**, 087801 (2006).

[47]  D. Salgado-Blanco, E. Diaz-Herrera, and C. I. Mendoza, Effect of the anchoring strength on the phase behaviour of discotic liquid crystals under face-on confinement, J. Phys.: Condens. Matter **31**, 105101 (2019).

[48]  G. Bautista-Carbajal, P. Gurin, S. Varga, and G. Odriozola, Phase diagram of hard squares in slit confinement, Sci. Rep. **8**, 8886 (2018).

[49]  M. Schoen, D. J. Diestler, and J. H. Cushman, Stratification induced order–disorder phase transitions in molecularly thin confined films, J. Chem. Phys. **101**, 6865 (1994).

[50]  T. Gruhn and M. Schoen, A grand canonical ensemble Monte Carlo study of confined planar and homeotropically anchored Gay-Berne films, Mol. Phys. **93**, 681 (1998).

[51]  J. K. Kim, F. Araoka, S. M. Jeong, S. Dhara, K. Ishikawa, and H. Takezoe, Bistable device using anchoring transition of nematic liquid crystals, Appl. Phys. Lett. **95**,

063505 (2009).

[52]   P. I. C. Teixeira, F. Barmes, C. Anquetil-Deck, and D. J. Cleaver, Simulation and theory of hybrid aligned liquid crystal films, Phys. Rev. E **79**, 011709 (2009).

[53]   R. Zwanzig, First-order phase transition in a gas of long thin rods, J. Chem. Phys. **39**, 1714 (1963).

[54]   S. Varga, Y. Martínez-Ratón, E. Velasco, G. Bautista-Carbajal, and G. Odriozola, Effect of orientational restriction on monolayers of hard ellipsoids, Phys. Chem. Chem. Phys. **18**, 4547 (2016).

[55]   F. Gámez, R. D. Acemel, and A. Cuetos, Demixing and nematic behaviour of oblate hard spherocylinders and hard spheres mixtures: Monte Carlo simulation and Parsons–Lee theory, Mol. Phys. **111**, 3136 (2013).

[56]   J. D. Parsons, Nematic ordering in a system of rods, Phys. Rev. A **19**, 1225 (1979).

[57]   S. D. Lee, A numerical investigation of nematic ordering based on a simple hard-rod model, J. Chem. Phys. **87**, 4972 (1987).

[58]   P. Poier, S. A. Egorov, C. N. Likos, and R. Blaak, Concentration-induced planar-to-homeotropic anchoring transition of stiff ring polymers on hard walls, Soft Matter **12**, 7983 (2016).

[59]   F. Barmes, and D. J. Cleaver, Computer simulation of a liquid-crystal anchoring transition, Phys. Rev. E **69**, 061705 (2004).